\documentclass[useAMS,usenatbib]{mnras}
\usepackage{graphics}  \usepackage{epsfig, color}  \usepackage{graphicx}
\usepackage[fleqn]{amsmath}  
\usepackage{epstopdf}

\def\ltsima{$\; \buildrel < \over \sim \;$}
\def\simlt{\lower.5ex\hbox{\ltsima}}  
\def\gtsima{$\; \buildrel >\over\sim\;$}  
\def\simgt{\lower.5ex\hbox{\gtsima}}  

\title[Stellar feedback from multiple sources]{Simulating feedback from nuclear clusters: the impact of multiple sources}

\author[Bourne \& Power]
       {Martin A. Bourne$^{1,2}$\thanks{mabourne@ast.cam.ac.uk} \&
         Chris Power$^3$\\
         $^1$Department of Physics \& Astronomy, University of Leicester,
         Leicester, LE1 7RH, UK\\
         $^{2}$Institute of Astronomy and Kavli Institute for Cosmology, University of Cambridge, Madingley Road, Cambridge, CB3 0HA, UK\\
         $^{3}$International Centre for Radio Astronomy Research,
         University of Western Australia, 35 Stirling Highway,
         Crawley,\\ Western Australia 6009, Australia}

\begin{document}

\date{}

\pagerange{\pageref{firstpage}--\pageref{lastpage}}
\pubyear{}

\maketitle

\label{firstpage}

\begin{abstract} Nuclear star clusters (NCs) are found to exist in the
centres of many galaxies and appear to follow scaling relations
similar to those of super-massive black holes. Previous analytical
work has suggested that such relations are a consequence of feedback
regulated growth. We explore this idea using high resolution
hydrodynamical simulations, focusing on the validity of the
simplifying assumptions made in analytical models. In particular, we
investigate feedback emanating from multiple stellar sources rather than from a single source, as is usually assumed, and show that collisions
betweens shells of gas swept up by feedback leads to momentum
cancellation and the formation of high density clumps and
filaments. This high density material is resistant both to expulsion
from the galaxy potential and to disruption by feedback; if it falls
back onto the NC, we expect the gas to be available for further star
formation or for feeding a central black hole. We also note our results
may have implications for the evolution of globular clusters and
stellar clusters in high redshift dark matter halos.
 \end{abstract}


\section{Introduction}

Nuclear star clusters (NCs) are commonly observed in the centres of a range of
galaxies; including late type spirals \citep{BokerEtal02, WalcherEtAl05},
early type spirals \citep{CarolloEtAl97} and dwarf ellipticals
\citep{CoteEtal06}. They have typical half light radii of a few parsecs
\citep{GehaEtAl02,BokerEtAl04, CoteEtal06} and masses of $10^{6}-10^{8}$
M$_{\odot}$ \citep{WalcherEtAl05}. Further, while late type galaxies tend only
to host a NC \citep{SatyapalEtAl09}, some more massive galaxies host both a
supermassive black hole (SMBH) and NC \citep{SethEtAl08}. In fact there
appears to be a dichotomy between NCs and SMBHs with the ratio $M_{\rm
  BH}/(M_{\rm BH} + M_{\rm NC})$ varying from $\sim 0$ in low mass galaxies
($M_{\rm sph}\simlt 10^{8}$ M$_{\odot}$) to $\sim 1$ in high mass galaxies
($M_{\rm sph}\simgt 10^{10}$ M$_{\odot}$) with some overlap in between
\citep{GrahamSpitler09}. 

Like SMBHs, NC properties are found to scale with the properties of
host galaxy bulge. \citet{FerrareseEtal06} observed that NC masses $M_{\rm NC}$
scale with bulge velocity dispersion $\sigma$ as,
\begin{equation}
M_{\rm NC}=10^{6.91}M_{\odot}\left(\frac{\sigma}{54\mbox{ km
    s$^{-1}$}}\right)^{4.27}.
\label{m_nc_sig_mass}
\end{equation}
The similarity of Eq~\ref{m_nc_sig_mass} to the $M_{\rm BH}-\sigma$
relation for SMBHs have led some \citep[e.g.][]{McLaughlinEtal06} to
speculate an origin driven by a common process \citep[e.g.][]{King05,MurrayEtal2005}
-- mass growth self-regulated by means of feedback. However, although this picture
is physically compelling for reasons we describe below, it is important
to note that it is (1) deduced from analytical arguments drawing on idealised
assumptions, which are likely to break down when more realistic assumptions are made
\citep[e.g.][]{NayakshinPower10,bourne14}, and (2) difficult to reconcile with recent
studies \citep[e.g.][]{Graham12, LeighEtAl12, ScottGraham13},
which suggest that the logarithmic slope of Eq~\ref{m_nc_sig_mass} may be shallower
($\sim 1.5-2.7$) than that found by \citet{FerrareseEtal06}.

In this Letter, we use hydrodynamical simulations to address (1) and
to establish the conditions under which the analytical solutions might
hold, deferring (2) to future work.
For example, \citet{McLaughlinEtal06} modeled a NC forming from a central gas
reservoir in the early stages of its host galaxy's evolution and regulating
its mass by
means of momentum-conserving feedback from stellar winds and supernovae
(cf. \S\ref{mom_fb_model}), which gives rise naturally to a scaling relation
between $M_{\rm NC}$ and $\sigma$ of the form given by equation
\ref{m_nc_sig_mass}. \citet{NayakshinEtal09b} showed that the dichotomy
between NCs and SMBHs can be explained if they co-evolve; in systems with
$\sigma\simlt 150$ km s$^{-1}$, the dynamical time is sufficiently
short compared to the Salpeter time, the characteristic timecale for
black hole growth, that gas accumulates and forms stars
preferentially, whereas in systems with $\sigma\simgt 150$ km
s$^{-1}$, the dynamical time exceeds the Salpeter time and so SMBH
growth is favoured over star formation. Although both of these analytical
studies offer an attractive explanation for the origin of NCs and their
relationship with SMBHs, they make important simplifying assumptions
that do not necessarily hold in more realistic circumstances.

We revisit \citet{McLaughlinEtal06} and test the validity of the
assumption that stellar feedback in the form of winds
and supernovae can be treated as emission from a single central point
source. Using hydrodynamical simulations of a {\it live} NC
embedded in the core of a dark matter halo, which we describe in
\S~\ref{simulations}, we compare and contrast results of feedback
from both single and multiple sources separately. If feedback is
momentum-driven, resolving it as emanating from multiple sources
rather than a single source results in a drop in its efficiency, to
the extent that it is not clear whether it could regulate NC growth, as we discuss in \S~\ref{results}. Finally, we discuss the
significance of this work in \S~\ref{discussion}, commenting on both
its applicability to star cluster formation in general and its implications for the growth of SMBHs in the early
Universe.

\section{Feedback from nuclear clusters}
\label{mom_fb_model}

Here we briefly outline the momentum-feedback model of \citet{McLaughlinEtal06} that we implement \citep[see also,][]{SilkRees98, King03,McQMcL12a}.
The momentum outflow rate arising from NC feedback is calculated as
a fraction of the Eddington rate $L_{Edd}/c$, i.e.
\begin{equation}
\dot{M}v_{w}=\lambda \frac{L_{Edd}}{c}=\lambda \frac{4\pi GM_{\rm CMO}}{\kappa}
\label{mom_flux}
\end{equation}
where $\dot{M}$ is the mass outflow rate from stars, $v_{w}$ is the wind
velocity, $\lambda$ is the fraction of momentum flux in terms of the
equivalent $L_{Edd}/c$, $L_{Edd}$ is the Eddington luminosity, $M_{\rm CMO}$
is the
mass of the central massive object (CMO) and
$\kappa =\sigma_{\rm T}/m_{\rm p}$ is the electron scattering
opacity. Taking into account the expected momentum output due to stellar winds
and SN II, \cite{McLaughlinEtal06} found $\lambda\sim 0.05$ for a
\citet{Chabrier2003} initial mass function (IMF).

The equation of motion of the shell of material swept up by an
outflow with momentum flux $\dot{M}v_{w}$ is 
\begin{equation}
\frac{d\left[M_{\rm sh}\left(R\right)\dot{R}\right]}{dt}=\dot{M}v_{w}-\frac{GM_{\rm sh}\left(R\right)[M_{\rm CMO}+M_{\rm
      Pot}(R)]}{R^{2}},
\label{EoM}
\end{equation}
where $\dot{M}v_{w}$ is given by equation \ref{mom_flux}, $R$ and $\dot{R}$ are the shell radius and radial velocity respectively,
$M_{\rm sh}(R)$ is the mass of the swept up shell at $R$ and is equal to the
mass of ambient gas originally enclosed within $R$ and $M_{\rm Pot}(R)$ is the
total mass of material remaining within $R$ i.e. any dark matter and stars.
Neglecting $M_{\rm CMO}$\footnote{This assumes that $M_{\rm CMO}<<M_{\rm Pot}$.} in the 2nd term on the RHS of equation \ref{EoM}, for an isothermal potential the terms
on the right hand side of equation \ref{EoM} balance when
\begin{equation}
M_{\rm CMO}=M_{\sigma}=\frac{f_{g}\kappa}{\lambda\pi G^{2}}\sigma^{4}
\label{m_sig_sis}
\end{equation} where $M_{\sigma}$ is the critical mass above which
feedback from the NC is able to drive clear the swept up shell of
material. Whilst for an isothermal profile one can find a simple
analytical solution involving the velocity dispersion of the halo
(which is a constant everywhere in this case) a more suitable halo
property for non-isothermal halos is the peak circular velocity,
$V_{\rm c, pk}=\sqrt{GM_{\rm DM}(R_{\rm pk})/R_{\rm pk}}$. In this
case, for large halo masses, $M_{\sigma}$ tends to \citep{McQMcL12a}:
\begin{equation}
M_{\sigma}\rightarrow\frac{f_{\rm g}\kappa}{\lambda\pi G^{2}}\frac{V_{\rm c,
    pk}^{4}}{4} = 9.12\times 10^{7}M_{\odot} \left(\frac{V_{\rm c, pk}}{200 \rm km
  s^{-1}}\right)^{4}\left(\frac{f_{\rm g}}{0.16}\right)\lambda^{-1},
\label{scale_rel}
\end{equation} 
which is analogous to equation \ref{m_sig_sis}. 

\section{Simulation Set-up}
\label{simulations}

Our simulations are run with the SPH code
Gadget-3, which has been modified to use the SPHS formulation \citep{ReadHayfield12}, as well
as the Wendland C2 kernel \citep{Wendland95, DehnenAly12} with 100
neighbours. Simulations are run in a static Hernquist {potential},
\begin{equation}
M_{\rm p}(r)=M_{200}\frac{r^2}{(r+a)^2}
\label{Menc_Hern}
\end{equation} where $M_{200}=10^{10}$ M$_{\odot}$ and $a=7.67$ kpc are the total mass and scale radius of the potential respectively. A gaseous
component modelled by $N=10^{6}$ SPH particles is also included,
extending from the centre of the halo to $r=4$ pc. The gas follows the
same profile as the halo potential with a gas fraction {$f_{\rm
g}=M_{\rm g}(r)/M_{\rm p}(r)=0.16$} and is initially in hydrostatic
equilibrium. The total gas mass is $M_{g}\simeq 435$ M$_{\odot}$ and hence the mass of an SPH particle is $m_{\rm SPH}=4.35\times10^{-4}$ M$_{\odot}$. 

{We perform two sets of simulations; the first consists of 250 star particles, set up in equilibrium with the background potential out to a radius or $4$ pc. The particles follow a Hernquist profile such that $M(r) = M_{\sigma}r^2/(r+4\mbox{ pc})^2$, where $M_{\sigma}=2.24\times 10^{6}$ M$_{\odot}$ is taken from equation \ref{scale_rel}. Given that we only model the cluster out to $r=4$ pc, the total mass of star particles in these runs is $\sim 0.25\times M_{\sigma}=5.6\times 10^{5}$ M$_{\odot}$ and hence each star particle has a mass of $2240$ M$_{\odot}$. The second set of runs are performed with a single star particle at the centre of the potential  with a mass of either $M_{\sigma}$ or $0.25\times M_{\sigma}$.}

Feedback {is produced by {\bf all}} star particles using a kernel weighted scheme similar
to the momentum feedback method of \citet{Costa14}. During a time step of length $\Delta t$ SPH particles
neighbouring a star particle receive a momentum kick of 
\begin{equation} 
p_{\rm inj, k} =\lambda\frac{L_{\rm Edd}}{c}\frac{m_{\rm SPH} W(r_{\rm
    k}-r_{\rm *}, h_{\rm *})}{\rho_{\rm g}(r_{\rm *})}\Delta t \;,
\label{mom_inj_rate}
\end{equation}
where $m_{\rm SPH}$ is the mass of an SPH particle, $W(r_{\rm k}-r_{\rm *},
h_{\rm *})$ is the kernel weight of a SPH particle relative to a star
particle, $h_{\rm *}$ is the star's {\it smoothing} length, calculated over
$100$ neighbours, and $\rho_{\rm g}(r_{\rm *})$ is the gas density at the
location of the star. Kicks are directed radially away from star particles. 

We perform all simulations with and without radiative cooling, which is
included using the optically thin cooling functions of \citet{KatzEtAl96} and \citet{Mashchenko08} above and below $10^{4}$ K respectively. While we find this has an effect for the multiple source runs, the impact is negligible in the single
source runs. Therefore, for simplicity, all results for single
source runs are without radiative cooling. We summarize the four simulations
presented in table \ref{tab:sims}. 

\begin{table} 
\centering 
\begin{tabular}{| c | c | c | c | c |} 
\hline  
Run & $N_{*}$ & $M_{*}$ & $m_{*}$ & cooling \\  
\hline 
S100N & $1$ & $2.24\times 10^{6}$ M$_{\odot}$ & $2.24\times 10^{6}$ M$_{\odot}$ & none \\ 
S025N & $1$ & $5.6\times 10^{5}$ M$_{\odot}$ & $5.6\times 10^{5}$ M$_{\odot}$ & none \\ 
M025N & $250$ & $5.6\times 10^{5}$ M$_{\odot}$ & $2240$ M$_{\odot}$ & none \\ 
M025C & $250$ & $5.6\times 10^{5}$ M$_{\odot}$ & $2240$ M$_{\odot}$ & included \\ 
\hline
\end{tabular} 
\caption{Summary of simulations showing (l-r) run name, number of {star particles} ($N_{*}$), total stellar mass ($M_{*}$){, star particle mass ($m_{*}$)} and radiative cooling. Run nomenclature takes the form {\it XYZ} where {\it X} defines whether there is a single (S) or multiple (M) feedback sources, {\it Y} defines the stellar mass used as a percentage of the $M_{\sigma}$ mass and {\it Z} defines whether radiative cooling is (C) or is not (N) included.}
\label{tab:sims} 
\end{table}

\section{Results}
\label{results}
\subsection{Feedback from a single source}

\begin{figure} 
\psfig{file=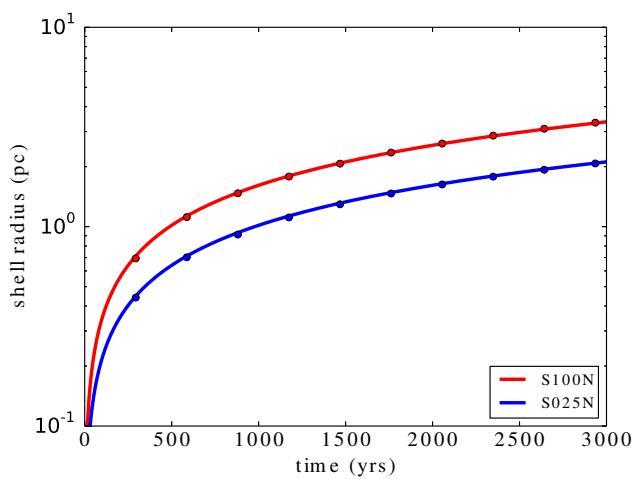,width=0.5\textwidth,angle=0}
\caption{Radius vs. time for the swept up shell of gas for a single
  feedback source with masses of $M_{\sigma}$ (red) and $0.25\times
  M_{\sigma}$ (blue). Solid points show results from simulations while lines show the analytical solution. The simulations match the analytical solution very well, vindicating our numerical method.}
\label{halo_revo} 
\end{figure}

As discussed in section \ref{mom_fb_model}, the evolution of a shell of gas
swept up by a single source of feedback can be calculated
analytically. Assuming a Hernquist background potential, as given in equation
\ref{Menc_Hern}, we numerically solve equation \ref{EoM} for a single central
source. For simplicity we assume that the gas also follows the Hernquist
profile with a gas fraction of $f_{\rm g}=0.16$. The results are shown by the
solid lines in Fig. \ref{halo_revo} where the the red and blue lines are for
CMO masses of $M_{\rm CMO}=M_{\sigma}$ and $M_{\sigma}/4$ respectively. In
both cases the shell of swept up gas is pushed outwards. 

We can compare the analytical solutions to the single source feedback
simulations that we have run. The solid points in Fig. \ref{halo_revo} show
the time evolution of the inner radius of the gas in the S100N and S025N
simulations. We can see that they match the analytical solution remarkably
well, suggesting that the feedback scheme we have implemented is able to
reproduce the analytical solution, at least in the simple, spherically
symmetric case. We note that the solution from the simulations also holds if
radiative cooling is included.

\subsection{Feedback from multiple sources}

\begin{figure*} 
\psfig{file=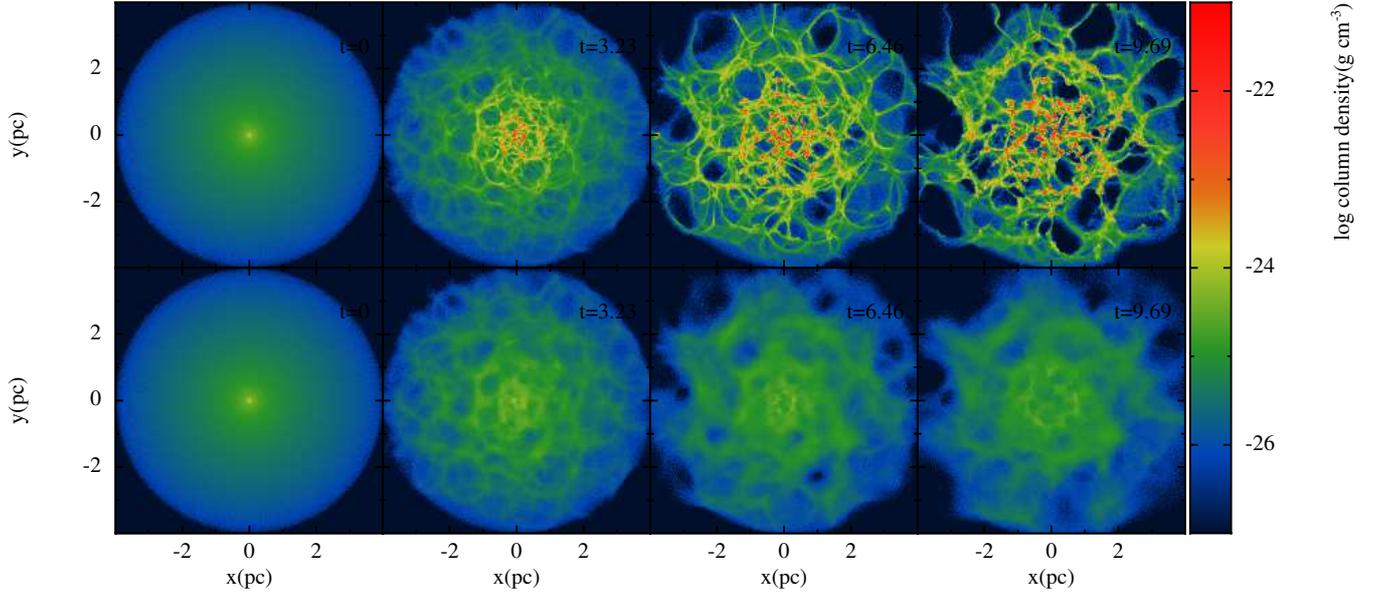,width=1.0\textwidth,angle=0}
\caption{Time evolution of the column density view of the M025C (top) and M025N (bottom) runs. These figures clearly illustrate the complex structure which forms when feedback is produced by multiple sources and also indicates the effect that radiative cooling has.}
\label{ColRhoEvo} 
\end{figure*}

Having validated our feedback scheme in the single source regime, we now
consider the effects of using multiple feedback sources. Each source emits
feedback at the rate described by equation \ref{mom_inj_rate} such that the
total momentum injected over the $r=4$ pc region is equal to that produced in
our single source runs with $M_{\rm CMO}=M_{\sigma}/4$. The effect of
including multiple feedback sources is evident in Fig. \ref{ColRhoEvo}, which
shows the evolution of gas column density with time -- from left to right,
at time intervals of $3.23$ kyrs, with the top and bottom rows showing runs with
and without radiative cooling respectively. 

In both cases a filamentary structure
forms as feedback from the multiple sources leads to shells of swept up gas
colliding and shocking against each other, resulting in the cancellation of momentum. These results are in stark contrast to the single source regime in which all of the gas is swept up into a well defined
outflowing shell.  

Also evident is that the inclusion
of radiative cooling has an impact on the exact properties of the filaments
that form. Gas that is able to cool and radiate away any thermalised energy produced in shocks is much more easily compressed and can reach higher
densities. As we show in the next section, this results in the gas being far
more resilient to expulsion by the feedback. 

\subsection{Single versus multiple sources}

\begin{figure} 
\psfig{file=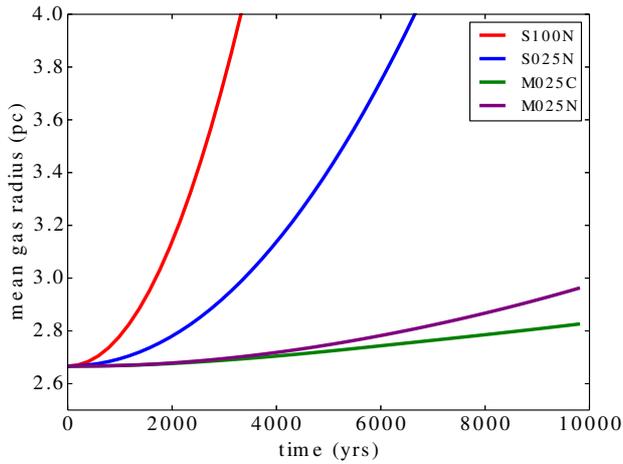,width=0.5\textwidth,angle=0}
\caption{Time evolution of the mean radial position of all gas particles for the single source runs with  CMO masses $M_{\sigma}$ (red) and $0.25\times
  M_{\sigma}$ (blue) and for multiple source runs with (green) and without (purple) cooling. It is clear that the single source runs clear out the gas on a much shorter timescale than the multiple source runs.}
\label{mean_r_t} 
\end{figure}

While it is possible to calculate the swept up shell radius in the single
feedback source runs, which thus provides a useful diagnostic, this is not
possible for the multiple source runs in which there is no well defined shell
of gas. This also means that there is no simple analytical solution with which
to compare the results of the multiple source runs, as we did for the single
source runs in Fig. \ref{halo_revo}. Therefore, as a proxy for the
effectiveness with which the feedback is able to {\it clear out} gas we
compare the mean radial position of the gas particles for all of the runs over
a $10$ kyr period. This is shown in Fig. \ref{mean_r_t}, where, as in Fig.
\ref{halo_revo}, the red and blue curves are for the S100N and S025N runs
respectively while the green and purple lines are for the M025C and M025N runs
respectively. 

The single source runs are clearly far more effective
at clearing out the gas on a short timescale whilst for the multiple source
runs the mean radius increases only slowly with time. We also note that
although similar initially, the mean radius for the M025N run becomes larger than
the M025C run at later times. As mentioned previously, radiative cooling
allows the gas in the M025C runs to be compressed to higher densities which
makes it far more resilient to the feedback and hence hard to push out. 

\begin{figure} 
\psfig{file=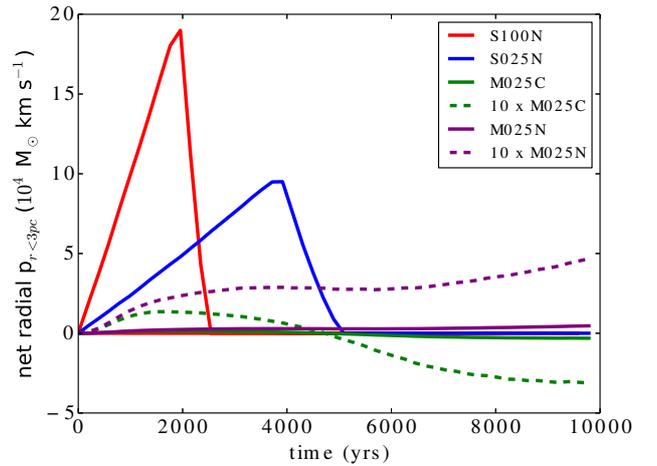,width=0.5\textwidth,angle=0}
\caption{Time evolution of the net gas radial momentum within 3pc for single source
  runs with CMO masses $M_{\sigma}$ (red) and $0.25\times
  M_{\sigma}$ (blue) and multiple source runs with (green) and without (purple) cooling. Due to large difference
  between the net radial momentum in the single and multiple source runs the
  dashed lines are included showing $10 \times$ the net radial
  momentum.}
\label{vr_t} 
\end{figure}

While Fig. \ref{mean_r_t} hints at the importance of radiative cooling
for the multiple source runs, a clearer diagnostic is to consider the
net radial momentum of the gas. In order to avoid particles with
unphysical velocities caused by boundary effects, in this analysis we
only include particles within the central 3 pc. This is plotted in
Fig. \ref{vr_t} and shows the difference between the multiple source
runs much more clearly. Firstly, however, let us compare the single
source runs (red and blue) with the multiple source runs (green and
purple). While the net radial momentum grows rapidly for the single
source runs, with the swept up shell of gas being accelerated out of
the centre of the galaxy, the net radial momentum for the multi-source
runs remains very small. Given this we also plot the net radial
momentum multiplied by a factor of $10$ for the multi-source runs
(dashed lines). The cancellation of momentum within the NC itself reduces the overall budget of momentum that could potentially escape the cluster, therefore, despite only simulating the central 4 pc, feedback at larger radii must also be reduced for multiple source runs.

Finally, focussing on the multi-source runs, what is also evident is that when
radiative cooling is included the net gas flow becomes negative, however when there
is no radiative cooling the gas, in general, is still swept up and pushed
out. This illustrates the importance of cooling when considering the
efficiency of feedback. 

\vspace{-1.5em}
\section{Discussion}
\label{discussion}
Simple analytical arguments have been used to suggest that NCs regulate
their growth by means of momentum-driven feedback -- winds and supernovae -- from
the stars that they form. In this letter we have assessed the validity of such
arguments by using hydrodynamical
simulations designed to explore the importance of the assumption that feedback
emanates from a single source. By studying the evolution of a shell of gas
within a dark matter halo and subjected to stellar feedback from a live star
cluster, we have found that feedback from multiple sources is less effective
than from a single source at sweeping up and clearing out gas from the central
regions of the halo. Feedback from multiple sources seed formation of
filamentary structures and high density clumps that are difficult to expel;
this effect is accentuated when ambient gas is allowed to cool. 

We found that the analytical model of \citet{McLaughlinEtal06} breaks down
  when more realistic assumptions are adopted and, consequently, it is unclear
  whether or not NCs and SMBHs should share similar $M_{\rm CMO}-\sigma$ relations.
  Indeed it may be that the observed difference between the power law index of
  $\sim 4$ for SMBHs and $\sim 2$ for NCs is a signature of a single source versus
  multiple sources of feedback, however this is beyond the scope of this work. We
  further highlight that our results suggest that the model developed by \citet{NayakshinEtal09b},
although physically compelling, needs revision. In that study, NCs grow at
the expense of SMBHs in low-mass galaxies by expelling gas efficiently that
would otherwise fuel the growth of an SMBH. However, in this study, the
efficiency of expulsion is reduced because clumps and filaments form in the
gas that are resistant to feedback; these fall back onto the NC and
either form new stars or potentially fuel the SMBH. Determining the ultimate
fate of these clumps and filaments will form the basis of future work. 

Our aim in this letter was to assess how simplifying assumptions made
in analytical studies of NC formation and evolution might affect the
validity of the predictions. However, we have made some simplifying
assumptions of our own, which we briefly consider here. First, as in
\citet{McLaughlinEtal06}, we combined all feedback into one continuous
injection of momentum at a fraction ($\lambda=0.05$) of the equivalent
Eddington momentum for the stellar population; in reality, the amount of
feedback released by a stellar population will be time dependent, with
stellar winds dominating the initial feedback phases and SNII becoming
more important at later times. This may have the effect of modifying or
enhancing the clumpiness and filamentarity within the gas distribution, as
a result of spatial inhomogeneities and time dependence of the feedback. We
will study this effect in future work that includes self-consistent
star formation and stellar lifetimes, although our intuition suggests that it
strengthens our conclusions. Second, we included 
only momentum-driven feedback in our models, but it is likely that
energy-driven feedback, especially due to SNII, will also play a role;
however, if an inhomogeneous and clumpy ambient medium is able to form, even
energy driven feedback can be ineffective at removing the gas from the central
regions of the galaxy \citep[e.g.,][]{bourne14}. {Finally we note that our results also complement understanding of globular cluster formation in which multiple stellar populations are expected to form during separate star formation episodes \citep[e.g.,][]{KrauseEtAl12, LeighEtAl13}.}

\vspace{-1.5em}
\section*{Acknowledgments}
We thank the referee for useful comments that helped to improve the clarity of this letter. MAB acknowledges an STFC grant and an STFC research studentship support. CP
acknowledges support of ARC Discovery Project DP140100198 and an ARC Future
Fellowship ARC FT130100041. This research used the DiRAC Complexity
system, operated by the University of Leicester IT Services, which forms part
of the STFC DiRAC HPC Facility (www.dirac.ac.uk). This equipment is funded by
BIS National E-Infrastructure capital grant ST/K000373/1 and  STFC DiRAC
Operations grant ST/K0003259/1. DiRAC is part of the UK National
E-Infrastructure. Figure \ref{ColRhoEvo} was produced
using SPLASH \citep{Price07}.

\vspace{-1.5em}
\bibliographystyle{mnras} \bibliography{nayakshin}

\begin{thebibliography}{}
\makeatletter
\relax
\def\mn@urlcharsother{\let\do\@makeother \do\$\do\&\do\#\do\^\do\_\do\%\do\~}
\def\mn@doi{\begingroup\mn@urlcharsother \@ifnextchar [ {\mn@doi@}
  {\mn@doi@[]}}
\def\mn@doi@[#1]#2{\def\@tempa{#1}\ifx\@tempa\@empty \href
  {http://dx.doi.org/#2} {doi:#2}\else \href {http://dx.doi.org/#2} {#1}\fi
  \endgroup}
\def\mn@eprint#1#2{\mn@eprint@#1:#2::\@nil}
\def\mn@eprint@arXiv#1{\href {http://arxiv.org/abs/#1} {{\tt arXiv:#1}}}
\def\mn@eprint@dblp#1{\href {http://dblp.uni-trier.de/rec/bibtex/#1.xml}
  {dblp:#1}}
\def\mn@eprint@#1:#2:#3:#4\@nil{\def\@tempa {#1}\def\@tempb {#2}\def\@tempc
  {#3}\ifx \@tempc \@empty \let \@tempc \@tempb \let \@tempb \@tempa \fi \ifx
  \@tempb \@empty \def\@tempb {arXiv}\fi \@ifundefined
  {mn@eprint@\@tempb}{\@tempb:\@tempc}{\expandafter \expandafter \csname
  mn@eprint@\@tempb\endcsname \expandafter{\@tempc}}}

\bibitem[\protect\citeauthoryear{{B{\"o}ker}, {Laine}, {van der Marel},
  {Sarzi}, {Rix}, {Ho}  \& {Shields}}{{B{\"o}ker} et~al.}{2002}]{BokerEtal02}
{B{\"o}ker} T.,  {Laine} S.,  {van der Marel} R.~P.,  {Sarzi} M.,  {Rix} H.-W.,
   {Ho} L.~C.,   {Shields} J.~C.,  2002, \mn@doi [\aj] {10.1086/339025}, \href
  {http://ukads.nottingham.ac.uk/abs/2002AJ....123.1389B} {123, 1389}

\bibitem[\protect\citeauthoryear{{B{\"o}ker}, {Sarzi}, {McLaughlin}, {van der
  Marel}, {Rix}, {Ho}  \& {Shields}}{{B{\"o}ker} et~al.}{2004}]{BokerEtAl04}
{B{\"o}ker} T.,  {Sarzi} M.,  {McLaughlin} D.~E.,  {van der Marel} R.~P.,
  {Rix} H.-W.,  {Ho} L.~C.,   {Shields} J.~C.,  2004, \mn@doi [\aj]
  {10.1086/380231}, \href {http://adsabs.harvard.edu/abs/2004AJ....127..105B}
  {127, 105}

\bibitem[\protect\citeauthoryear{{Bourne}, {Nayakshin}  \& {Hobbs}}{{Bourne}
  et~al.}{2014}]{bourne14}
{Bourne} M.~A.,  {Nayakshin} S.,   {Hobbs} A.,  2014, \mn@doi [\mnras]
  {10.1093/mnras/stu747}, \href
  {http://adsabs.harvard.edu/abs/2014MNRAS.441.3055B} {441, 3055}

\bibitem[\protect\citeauthoryear{{Carollo}, {Stiavelli}, {de Zeeuw}  \&
  {Mack}}{{Carollo} et~al.}{1997}]{CarolloEtAl97}
{Carollo} C.~M.,  {Stiavelli} M.,  {de Zeeuw} P.~T.,   {Mack} J.,  1997,
  \mn@doi [\aj] {10.1086/118654}, \href
  {http://adsabs.harvard.edu/abs/1997AJ....114.2366C} {114, 2366}

\bibitem[\protect\citeauthoryear{{Chabrier}}{{Chabrier}}{2003}]{Chabrier2003}
{Chabrier} G.,  2003, \mn@doi [\pasp] {10.1086/376392}, \href
  {http://adsabs.harvard.edu/abs/2003PASP..115..763C} {115, 763}

\bibitem[\protect\citeauthoryear{{Costa}, {Sijacki}  \& {Haehnelt}}{{Costa}
  et~al.}{2014}]{Costa14}
{Costa} T.,  {Sijacki} D.,   {Haehnelt} M.~G.,  2014, \mn@doi [\mnras]
  {10.1093/mnras/stu1632}, \href
  {http://adsabs.harvard.edu/abs/2014MNRAS.444.2355C} {444, 2355}

\bibitem[\protect\citeauthoryear{{C{\^o}t{\'e}} et~al.,}{{C{\^o}t{\'e}}
  et~al.}{2006}]{CoteEtal06}
{C{\^o}t{\'e}} P.,  et~al., 2006, \mn@doi [\apjs] {10.1086/504042}, \href
  {http://ukads.nottingham.ac.uk/abs/2006ApJS..165...57C} {165, 57}

\bibitem[\protect\citeauthoryear{{Dehnen} \& {Aly}}{{Dehnen} \&
  {Aly}}{2012}]{DehnenAly12}
{Dehnen} W.,  {Aly} H.,  2012, \mn@doi [\mnras]
  {10.1111/j.1365-2966.2012.21439.x}, \href
  {http://adsabs.harvard.edu/abs/2012MNRAS.425.1068D} {425, 1068}

\bibitem[\protect\citeauthoryear{{Ferrarese} et~al.,}{{Ferrarese}
  et~al.}{2006}]{FerrareseEtal06}
{Ferrarese} L.,  et~al., 2006, \mn@doi [\apjl] {10.1086/505388}, \href
  {http://ukads.nottingham.ac.uk/cgi-bin/nph-bib_query?bibcode=2006ApJ...644L..21F&db_key=AST}
  {644, L21}

\bibitem[\protect\citeauthoryear{{Geha}, {Guhathakurta}  \& {van der
  Marel}}{{Geha} et~al.}{2002}]{GehaEtAl02}
{Geha} M.,  {Guhathakurta} P.,   {van der Marel} R.~P.,  2002, \mn@doi [\aj]
  {10.1086/344764}, \href {http://adsabs.harvard.edu/abs/2002AJ....124.3073G}
  {124, 3073}

\bibitem[\protect\citeauthoryear{{Graham}}{{Graham}}{2012}]{Graham12}
{Graham} A.~W.,  2012, \mn@doi [\mnras] {10.1111/j.1365-2966.2012.20734.x},
  \href {http://adsabs.harvard.edu/abs/2012MNRAS.422.1586G} {422, 1586}

\bibitem[\protect\citeauthoryear{{Graham} \& {Spitler}}{{Graham} \&
  {Spitler}}{2009}]{GrahamSpitler09}
{Graham} A.~W.,  {Spitler} L.~R.,  2009, \mn@doi [\mnras]
  {10.1111/j.1365-2966.2009.15118.x}, \href
  {http://adsabs.harvard.edu/abs/2009MNRAS.397.2148G} {397, 2148}

\bibitem[\protect\citeauthoryear{{Katz}, {Weinberg}  \& {Hernquist}}{{Katz}
  et~al.}{1996}]{KatzEtAl96}
{Katz} N.,  {Weinberg} D.~H.,   {Hernquist} L.,  1996, \mn@doi [\apjs]
  {10.1086/192305}, \href {http://adsabs.harvard.edu/abs/1996ApJS..105...19K}
  {105, 19}

\bibitem[\protect\citeauthoryear{{King}}{{King}}{2003}]{King03}
{King} A.,  2003, \mn@doi [\apjl] {10.1086/379143}, \href
  {http://adsabs.harvard.edu/cgi-bin/nph-bib_query?bibcode=2003ApJ...596L..27K&db_key=AST}
  {596, L27}

\bibitem[\protect\citeauthoryear{{King}}{{King}}{2005}]{King05}
{King} A.,  2005, \mn@doi [\apjl] {10.1086/499430}, \href
  {http://adsabs.harvard.edu/cgi-bin/nph-bib_query?bibcode=2005ApJ...635L.121K&db_key=AST}
  {635, L121}

\bibitem[\protect\citeauthoryear{{Krause}, {Charbonnel}, {Decressin}, {Meynet},
  {Prantzos}  \& {Diehl}}{{Krause} et~al.}{2012}]{KrauseEtAl12}
{Krause} M.,  {Charbonnel} C.,  {Decressin} T.,  {Meynet} G.,  {Prantzos} N.,
  {Diehl} R.,  2012, \mn@doi [\aap] {10.1051/0004-6361/201220244}, \href
  {http://adsabs.harvard.edu/abs/2012A%26A...546L...5K} {546, L5}

\bibitem[\protect\citeauthoryear{{Leigh}, {B{\"o}ker}  \& {Knigge}}{{Leigh}
  et~al.}{2012}]{LeighEtAl12}
{Leigh} N.,  {B{\"o}ker} T.,   {Knigge} C.,  2012, \mn@doi [\mnras]
  {10.1111/j.1365-2966.2012.21365.x}, \href
  {http://adsabs.harvard.edu/abs/2012MNRAS.424.2130L} {424, 2130}

\bibitem[\protect\citeauthoryear{{Leigh}, {B{\"o}ker}, {Maccarone}  \&
  {Perets}}{{Leigh} et~al.}{2013}]{LeighEtAl13}
{Leigh} N.~W.~C.,  {B{\"o}ker} T.,  {Maccarone} T.~J.,   {Perets} H.~B.,  2013,
  \mn@doi [\mnras] {10.1093/mnras/sts554}, \href
  {http://adsabs.harvard.edu/abs/2013MNRAS.429.2997L} {429, 2997}

\bibitem[\protect\citeauthoryear{{Mashchenko}, {Wadsley}  \&
  {Couchman}}{{Mashchenko} et~al.}{2008}]{Mashchenko08}
{Mashchenko} S.,  {Wadsley} J.,   {Couchman} H.~M.~P.,  2008, \mn@doi [Science]
  {10.1126/science.1148666}, \href
  {http://adsabs.harvard.edu/abs/2008Sci...319..174M} {319, 174}

\bibitem[\protect\citeauthoryear{{McLaughlin}, {King}  \&
  {Nayakshin}}{{McLaughlin} et~al.}{2006}]{McLaughlinEtal06}
{McLaughlin} D.~E.,  {King} A.~R.,   {Nayakshin} S.,  2006, \mn@doi [\apjl]
  {10.1086/508627}, \href
  {http://ukads.nottingham.ac.uk/abs/2006ApJ...650L..37M} {650, L37}

\bibitem[\protect\citeauthoryear{{McQuillin} \& {McLaughlin}}{{McQuillin} \&
  {McLaughlin}}{2012}]{McQMcL12a}
{McQuillin} R.~C.,  {McLaughlin} D.~E.,  2012, \mn@doi [\mnras]
  {10.1111/j.1365-2966.2012.21028.x}, \href
  {http://adsabs.harvard.edu/abs/2012MNRAS.423.2162M} {423, 2162}

\bibitem[\protect\citeauthoryear{{Murray}, {Quataert}  \& {Thompson}}{{Murray}
  et~al.}{2005}]{MurrayEtal2005}
{Murray} N.,  {Quataert} E.,   {Thompson} T.~A.,  2005, \mn@doi [\apj]
  {10.1086/426067}, \href {http://adsabs.harvard.edu/abs/2005ApJ...618..569M}
  {618, 569}

\bibitem[\protect\citeauthoryear{{Nayakshin} \& {Power}}{{Nayakshin} \&
  {Power}}{2010}]{NayakshinPower10}
{Nayakshin} S.,  {Power} C.,  2010, \mn@doi [\mnras]
  {10.1111/j.1365-2966.2009.15946.x}, \href
  {http://adsabs.harvard.edu/abs/2010MNRAS.402..789N} {402, 789}

\bibitem[\protect\citeauthoryear{{Nayakshin}, {Wilkinson}  \&
  {King}}{{Nayakshin} et~al.}{2009}]{NayakshinEtal09b}
{Nayakshin} S.,  {Wilkinson} M.~I.,   {King} A.,  2009, \mn@doi [\mnras]
  {10.1111/j.1745-3933.2009.00709.x}, \href
  {http://ukads.nottingham.ac.uk/abs/2009MNRAS.398L..54N} {398, L54}

\bibitem[\protect\citeauthoryear{{Price}}{{Price}}{2007}]{Price07}
{Price} D.~J.,  2007, \mn@doi [\pasa] {10.1071/AS07022}, \href
  {http://adsabs.harvard.edu/abs/2007PASA...24..159P} {24, 159}

\bibitem[\protect\citeauthoryear{{Read} \& {Hayfield}}{{Read} \&
  {Hayfield}}{2012}]{ReadHayfield12}
{Read} J.~I.,  {Hayfield} T.,  2012, \mn@doi [\mnras]
  {10.1111/j.1365-2966.2012.20819.x}, \href
  {http://adsabs.harvard.edu/abs/2012MNRAS.422.3037R} {422, 3037}

\bibitem[\protect\citeauthoryear{{Satyapal}, {B{\"o}ker}, {Mcalpine},
  {Gliozzi}, {Abel}  \& {Heckman}}{{Satyapal} et~al.}{2009}]{SatyapalEtAl09}
{Satyapal} S.,  {B{\"o}ker} T.,  {Mcalpine} W.,  {Gliozzi} M.,  {Abel} N.~P.,
  {Heckman} T.,  2009, \mn@doi [\apj] {10.1088/0004-637X/704/1/439}, \href
  {http://adsabs.harvard.edu/abs/2009ApJ...704..439S} {704, 439}

\bibitem[\protect\citeauthoryear{{Scott} \& {Graham}}{{Scott} \&
  {Graham}}{2013}]{ScottGraham13}
{Scott} N.,  {Graham} A.~W.,  2013, \mn@doi [\apj]
  {10.1088/0004-637X/763/2/76}, \href
  {http://adsabs.harvard.edu/abs/2013ApJ...763...76S} {763, 76}

\bibitem[\protect\citeauthoryear{{Seth}, {Ag{\"u}eros}, {Lee}  \&
  {Basu-Zych}}{{Seth} et~al.}{2008}]{SethEtAl08}
{Seth} A.,  {Ag{\"u}eros} M.,  {Lee} D.,   {Basu-Zych} A.,  2008, \mn@doi
  [\apj] {10.1086/528955}, \href
  {http://ukads.nottingham.ac.uk/abs/2008ApJ...678..116S} {678, 116}

\bibitem[\protect\citeauthoryear{{Silk} \& {Rees}}{{Silk} \&
  {Rees}}{1998}]{SilkRees98}
{Silk} J.,  {Rees} M.~J.,  1998, \aap, \href
  {http://adsabs.harvard.edu/abs/1998A%26A...331L...1S} {331, L1}

\bibitem[\protect\citeauthoryear{{Walcher} et~al.,}{{Walcher}
  et~al.}{2005}]{WalcherEtAl05}
{Walcher} C.~J.,  et~al., 2005, \mn@doi [\apj] {10.1086/425977}, \href
  {http://ukads.nottingham.ac.uk/abs/2005ApJ...618..237W} {618, 237}

\bibitem[\protect\citeauthoryear{Wendland}{Wendland}{1995}]{Wendland95}
Wendland H.,  1995, Advances in computational Mathematics, 4, 389

\makeatother
\end{thebibliography}
\label{lastpage}

\end{document}